\documentclass[sigconf, screen]{acmart}

\usepackage{url}

\usepackage{bm}
\usepackage{amsmath}

\usepackage{amssymb}

\usepackage{graphicx}
\usepackage{textcomp}
\usepackage{color, colortbl}
\usepackage{xcolor}
\usepackage{caption}

\usepackage{tcolorbox}

\usepackage{multirow}
\usepackage[noend]{algpseudocode}
\usepackage{algorithm}
\usepackage{xspace}
\usepackage{tikz}
\usepackage{pifont}
\usepackage{booktabs}

\makeatletter
\DeclareRobustCommand\onedot{\futurelet\@let@token\@onedot}
\def\@onedot{\ifx\@let@token.\else.\null\fi\xspace}
\def\eg{\emph{e.g}\onedot}

\def\etc{\emph{etc}\onedot} 
 
\def\etal{\emph{et al}\onedot}
\makeatother

\definecolor{dg}{rgb}{0,0.694,0.298}
\definecolor{purple}{rgb}{0.4,0.176,0.569}
\definecolor{iris}{rgb}{0.35, 0.31, 0.81}
\definecolor{tabgray}{rgb}{0.85,0.85,0.85}

\newcommand{\zhijie}[1]{\textbf{\textcolor{purple}{Zhijie: #1}}}

\usepackage{todonotes}

\newcommand{\myparagraph}[1]{\vspace{0.5em}\noindent{\bf #1.}}

\def\BibTeX{{\rm B\kern-.05em{\sc i\kern-.025em b}\kern-.08em
    T\kern-.1667em\lower.7ex\hbox{E}\kern-.125emX}}

\newcommand{\globalBudget}{\ensuremath{B_G}\xspace}
\newcommand{\rb}{\ensuremath{\mathsf{rb}}\xspace}

\newcommand{\Robust}[2]{{ \mathsf{Rob}(#1, #2) }}
\newcommand{\dimension}{\ensuremath{\mathsf{dim}}\xspace}

\usepackage{subcaption}
\usepackage{paralist}
\usepackage{tabularx}
\usepackage{enumitem}
\newcommand{\UntilOp}[1]{\mathbin{\mathcal{U}_{#1}}}
\newcommand{\matlab}{\textsc{Matlab}\xspace}

\newcommand{\vego}{\ensuremath{v_{\mathit{ego}}}\xspace}

\newcommand{\vset}{\ensuremath{v_{\mathit{set}}}\xspace}

\newcommand{\drel}{\ensuremath{d_{\mathit{rel}}}\xspace}

\newcommand{\dsafe}{\ensuremath{d_{\mathit{safe}}}\xspace}

\newcommand{\pedal}{\ensuremath{\mathit{PedalAngle}}\xspace}
\newcommand{\engine}{\ensuremath{\mathit{EngineSpeed}}\xspace}
\newcommand{\spec}[1]{\ensuremath{\mathsf{S}#1}\xspace}
\newcommand{\boldspec}[1]{\ensuremath{\boldsymbol{\mathsf{S}}\mathbf{#1}}\xspace}

\newcommand{\af}{\ensuremath{\mathit{AF}}\xspace}
\newcommand{\afref}{\ensuremath{\mathit{AFref}}\xspace}
\newcommand{\fs}{\ensuremath{\mathit{F}_{\mathit{s}}}\xspace}
\newcommand{\pref}{\ensuremath{\mathit{P}_{\mathit{ref}}}\xspace}

\newcommand{\nr}{\ensuremath{\mathit{n_r}}\xspace}
\newcommand{\nv}{\ensuremath{\mathit{n_v}}\xspace}

\newcommand{\acc}{\texttt{ACC}\xspace}
\newcommand{\lka}{\texttt{LKA}\xspace}
\newcommand{\ddpg}{\texttt{DDPG}\xspace}
\newcommand{\tdthr}{\texttt{TD3}\xspace}
\newcommand{\sac}{\texttt{SAC}\xspace}
\newcommand{\ppo}{\texttt{PPO}\xspace}
\newcommand{\atwoc}{\texttt{A2C}\xspace}
\newcommand{\trad}{\texttt{T}\xspace}
\newcommand{\hb}{\texttt{HS}\xspace}
\newcommand{\hbra}{\texttt{HR\_0.1}\xspace}
\newcommand{\hbrb}{\texttt{HR\_1}\xspace}
\newcommand{\hba}{\texttt{HA}\xspace}

\newcommand{\afc}{\texttt{AFC}\xspace}
\newcommand{\wt}{\texttt{WT}\xspace}
\newcommand{\scd}{\texttt{SC}\xspace}
\newcommand{\wtk}{\texttt{WTK}\xspace}
\newcommand{\lr}{\texttt{LR}\xspace}
\newcommand{\cstr}{\texttt{CSTR}\xspace}
\newcommand{\apv}{\texttt{APV}\xspace}

\AtBeginDocument{%
  \providecommand\BibTeX{{%
    \normalfont B\kern-0.5em{\scshape i\kern-0.25em b}\kern-0.8em\TeX}}}

\copyrightyear{2022}
\acmYear{2022}
\setcopyright{acmcopyright}\acmConference[ICSE-SEIP '22]{44nd International Conference on Software Engineering: Software Engineering in Practice}{May 21--29, 2022}{Pittsburgh, PA, USA}
\acmBooktitle{44nd International Conference on Software Engineering: Software Engineering in Practice (ICSE-SEIP '22), May 21--29, 2022, Pittsburgh, PA, USA}
\acmPrice{15.00}
\acmDOI{10.1145/3510457.3513049}
\acmISBN{978-1-4503-9226-6/22/05}

\begin{document}

\title[When Cyber-Physical Systems Meet AI: A Benchmark, an Evaluation, and a Way Forward]{When Cyber-Physical Systems Meet AI: A Benchmark, an Evaluation, and a Way Forward}

\author{Jiayang Song$^{1\dagger}$, Deyun Lyu$^{2\dagger}$, Zhenya Zhang$^3$, Zhijie Wang$^1$, Tianyi Zhang$^4$, Lei Ma$^{1,2,5}$}
\affiliation{%
  \institution{$^1${University of Alberta, Canada}\quad $^2${Kyushu University, Japan}  \quad$^3${Nanyang Technological University, Singapore}\\ \quad$^4${Purdue University, USA}
  \quad$^5${Alberta Machine Intelligence Institute (Amii), Canada}\\
  jiayan13@ualberta.ca, lyu.deyun.107@s.kyushu-u.ac.jp, zhenya.zhang@ntu.edu.sg,\\zhijie.wang@ualberta.ca, tianyi@purdue.edu, ma.lei@acm.org}
\country{}
 }

\settopmatter{authorsperrow=1} 


\thanks{$^{\mathrm{\dagger}}$ Both authors contributed equally to this research.}

\renewcommand{\shortauthors}{J. Song, D. Lyu, Z. Zhang, Z. Wang, T. Zhang, and L. Ma}

\begin{abstract}
Cyber-Physical Systems (CPS) have been broadly deployed in safety-critical domains, such as automotive systems, avionics, medical devices,  \etc. In recent years, Artificial Intelligence (AI) has been increasingly adopted to control CPS. Despite the popularity of AI-enabled CPS, few benchmarks are publicly available. There is also a lack of deep understanding on the performance and reliability of AI-enabled CPS across different industrial domains.
To bridge this gap, we present a public benchmark of industry-level CPS in seven domains and build AI controllers for them via state-of-the-art \emph{deep reinforcement learning (DRL)} methods. Based on that, we further perform a systematic evaluation of these AI-enabled systems with their traditional counterparts to identify current challenges and future opportunities. Our key findings include (1) AI controllers do not always outperform traditional controllers, (2) existing CPS testing techniques (falsification, specifically) fall short of analyzing AI-enabled CPS, and (3) building a hybrid system that strategically combines and switches between AI controllers and traditional controllers can achieve better performance across different domains. Our results highlight the need for new testing techniques for AI-enabled CPS and the need for more investigations into hybrid CPS to achieve optimal performance and reliability. 
Our benchmark, code, detailed evaluation results, and experiment scripts are available on \href {https://sites.google.com/view/ai-cps-benchmark}{https://sites.google.com/view/ai-cps-benchmark}.

\end{abstract}



\keywords{Cyper-Physical Systems, AI Controllers, Falsification, Benchmarks, Deep Reinforcement Learning}

\maketitle

\section{Introduction}

\begin{figure*}
\centering
\includegraphics[width=\linewidth]{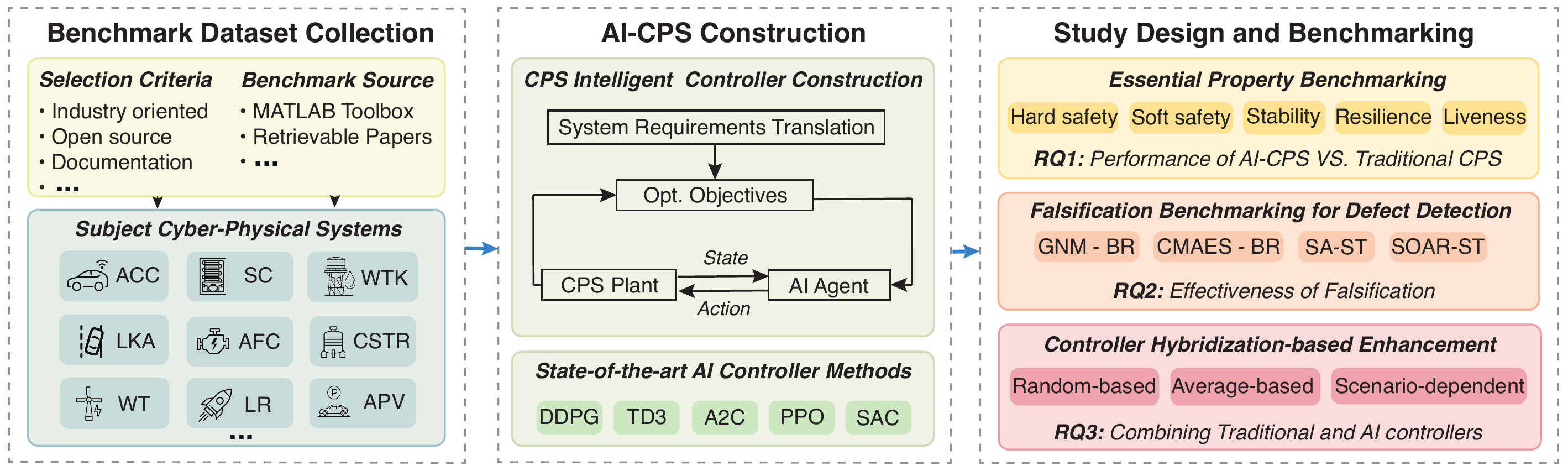}
\caption{Workflow summary of AI-enabled CPS dataset and benchmark construction, and high-level empirical study design.}
\vspace{-10pt}
\label{figs:workflow}
\end{figure*}

Cyber-Physical Systems (CPS) refer to the combination of mechanical and computer systems, in which computer systems actively monitor and control the behavior of mechanical systems according to system states and external environments. The integration of computer systems significantly improves the performance of mechanical systems. Nowadays, CPS have been widely deployed in diverse industrial domains, such as automotive systems, energy control, avionics, medical devices, \etc.

With recent advances in Artificial Intelligence (AI), there has been increasing demand, from both industry and academia, in enhancing or even replacing traditional controllers with AI controllers (\eg, deep neural networks), in order to achieve more optimized and flexible control. Given the great learning and generalization capabilities of deep neural networks, AI controllers have been increasingly deployed in various domains to handle complex situations in the physical world~\cite{liu2021recurrent, nivison2017development}.

Despite the rapid development of AI-enabled CPS, there is a lack of comprehensive analysis of their performance (\eg, safety, reliability, robustness) in different domains. The main reason is that unlike the large and growing open-source community in conventional software, CPS are often kept as private intellectual properties by industrial practitioners, in which strong domain knowledge and know-how are encoded. Therefore, even up to the present, very few benchmarks of AI-enabled CPS are publicly available for such analysis. Furthermore, since existing quality assurance methods such as \emph{falsification}~\cite{donze2010breach, annpureddy2011s, zhang2018two, dreossi2019compositional, yamagata2020falsification, adimoolam2017classification} are mostly designed for traditional CPS controllers, 
it is unclear to what extent these methods are still effective in analyzing AI controllers. 

To bridge this gap, in this paper, we take the first step of benchmarking AI-enabled CPS in multiple industrial domains and performing a systematic analysis of their performance in comparison to their traditional counterparts. Fig.~\ref{figs:workflow} shows our high-level study workflow. In particular, we  
investigate three main research questions to identify the challenges and potential opportunities for building safe and reliable AI-enabled CPS:
\begin{compactitem}[$\bullet$]
\item \textbf{RQ1. How well do the DRL-based AI controllers perform compared with the traditional controllers?} 
This RQ aims to establish a comprehensive understanding of the advantages and limitations of both traditional and AI controllers in CPS. Our experiment shows that AI controllers do not always outperform their traditional counterparts. In several cases, AI controllers have weaknesses in handling multiple control outputs and fail to balance among multiple requirements.
\item \textbf{RQ2: To what extent are existing CPS testing methods still effective on AI-based CPS?} Falsification is a commonly used technique to detect defects in traditional CPS. Although many falsification techniques have been proposed, it is still unclear whether they are still effective in the context of AI-enabled CPS. This RQ aims to establish a testing benchmark of different falsification methods on both traditional and AI-enabled CPS. Our comparative study finds that existing falsification methods are mostly designed for traditional controllers, and are not effective enough for AI-enabled CPS. 
\item \textbf{RQ3: Can the combination of traditional and AI controllers bring better performance?} In practice, building a hybrid controller that strategically switches between AI controllers and traditional controllers can be a promising direction for better performance (\eg, similar to component redundancies for safety in ISO 26262~\cite{iso26262}). This RQ aims to make an early investigation on whether this can be a promising direction in the context of AI-enabled CPS. Overall, we find that among the three types of hybrid controllers we explored, a scenario-dependent approach outperforms the other two in most of the cases. This result confirms that strategically combining traditional and AI controllers is a promising direction for further research.
\end{compactitem}

In summary, this work makes the following contributions:
\begin{compactitem}[$\bullet$]
\item We create the first public dataset and benchmark of AI-enabled CPS that span over various industrial domains, such as driving assistants, chemical reactor, aerospace, powertrains, \etc. This provides a common ground for evaluating AI-based CPS and also enables further research along this direction.
\item We perform a systematic analysis of the performance (\eg, safety, reliability, robustness) of AI-enabled CPS, and benchmarks the current techniques (\eg, falsification, system enhancement) as the basis for further investigation.
\item Based on the analysis results, we further pose discussions on the future directions of AI-enabled CPS, including effectiveness of AI controllers, testing tools for AI-enabled CPS, and methods to construct hybrid control systems.
\end{compactitem}
To the best of our knowledge, this is the very first paper that establishes a publicly available dataset and benchmark for industry-level CPS with AI controllers. The benchmark and our empirical study results demonstrate the potential research opportunities around AI-enabled CPS to meet the growing industrial demands. Our work enables better understanding, establishes the basis, and paves the path towards 
further quality assurance research to build safe and reliable AI-enabled CPS.

\section{Background}

This section gives an introduction on CPS and AI controllers for CPS. Specifically, we describe deep reinforcement learning (DRL) based AI controllers in this paper. We also briefly describe \emph{Signal Temporal Logic (STL)}, a specification language of CPS, and \emph{falsification}, a typical and important testing method for CPS based on STL.

\subsection{Cyber-Physical Systems and AI controllers}

\begin{figure}[!tb]
\centering
\begin{subfigure}[b]{0.49\columnwidth}
\includegraphics[width=\columnwidth]{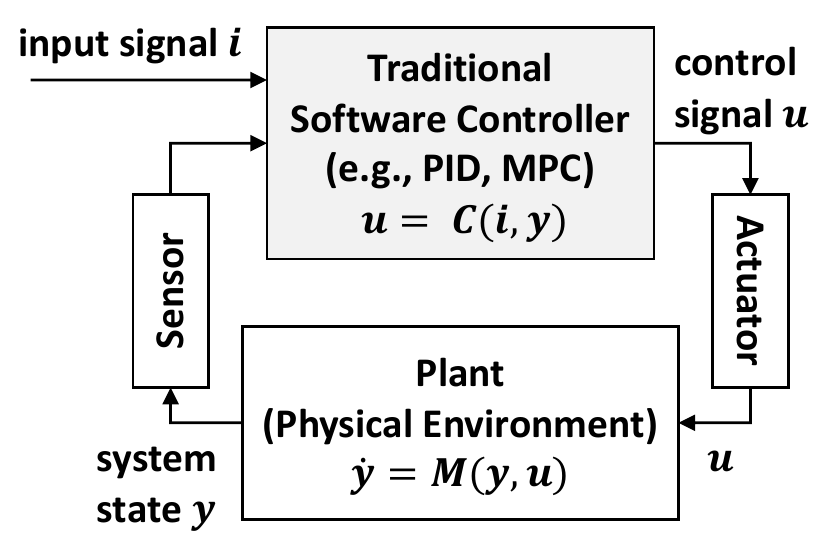}
\caption{Traditional CPS}
\label{fig:CPStradController}
\end{subfigure}
\hfill
\begin{subfigure}[b]{0.49\columnwidth}
\includegraphics[width=\columnwidth]{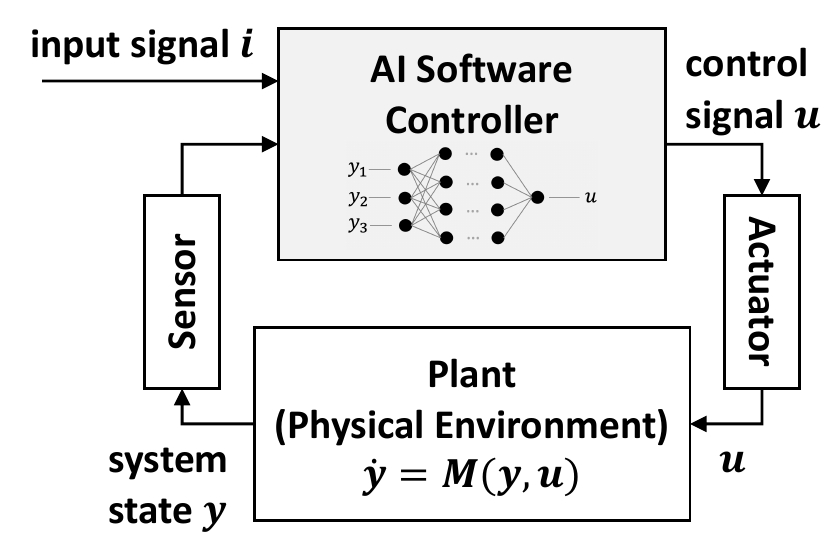}
\caption{AI-enabled CPS}
\label{fig:CPSAIController}
\end{subfigure}
\caption{Abstract workflow of CPS, with a traditional controller and an AI controller, respectively}
\end{figure}

CPS have been widely adopted in safety-critical industrial domains, such as automotive systems, avionics, medical devices, \etc. CPS make use of computer programs to monitor and control the behaviors of mechanical systems. 
Fig.~\ref{fig:CPStradController} shows a brief overview of a CPS, which consists of a plant $M$ and a controller $C$. The plant is a physical environment, whose next state $y$ is decided by the current state and the control command $u$. The control command $u$ is issued by a software controller. Commonly-used traditional controllers include \emph{proportional integral derivative (PID) control}, \emph{model predictive control (MPC)}, \etc~(detailed in  \S{}\ref{sec:benchmarks}). These controllers make the control decision $u$ based on the state $y$ and an external input $i$.  
 
In CPS, \emph{signals} are passed between components in a system and between the system and the external environment. Formally, a signal $s\colon~ [0, T] \to \mathbb{R}^{\dimension}$ is defined as a \emph{time-variant} function, where $T\in \mathbb{R}_{+}$ denotes the time horizon, and $\dimension\in \mathbb{N}_{+}$ is called the dimension of the signal. In Fig.~\ref{fig:CPStradController}, the system state $y$, the control decision $u$, and the external input $i$ are all signals. Sensors and actuators are used to collect, process, and pass these signals between the plant $M$ and the controller $C$ in a CPS model. 

\myparagraph{AI Controller} With the boom of AI in the past few years, practitioners have started considering replacing traditional controllers with AI controllers. Fig.~\ref{fig:CPSAIController} gives an overview of  \emph{AI-enabled CPS}. 
An AI controller can be implemented and trained in different ways, such as \emph{supervised learning} and \emph{reinforcement learning}. Among different methods, deep reinforcement learning (DRL) is considered as the state-of-the-art, and has succeeded in various application domains~\cite{arulkumaran2017deep}. In this study, we mainly focus on DRL-based controllers. We briefly introduce DRL below.
 
\myparagraph{Deep Reinforcement Learning} Unlike supervised learning, DRL does not require collecting labeled datasets beforehand. Instead, it follows the \emph{``trial-and-error''} paradigm---learning the best strategy via repeated interactions with the external environment. 
Technically, DRL requires to train an \emph{agent} in an environment so that the agent learns the best \emph{policy} that maximizes the cumulative \emph{reward}. At each step, the agent has a \emph{state} given by the environment and the agent itself, and it is required to take an \emph{action} to transit from the current state to a new one. A policy is a function that maps a state-action pair to a real-valued reward~\cite{balakrishnan2019structured}. For a fixed state, the reward of each action depends on the feedback from the environment. In order to learn the best policy, the agent repeatedly tries an action to obtain a reward, and updates the policy accordingly. There is a variety of policy updating methods, such as \emph{Deep deterministic policy gradient (DDPG)}~\cite{lillicrap2015continuous}, \emph{Twin-delayed deep deterministic policy gradient (TD3)}~\cite{fujimoto2018addressing}, \emph{Actor-critic (A2C)}~\cite{donze2010robust}, \emph{Proximal policy optimization (PPO)}~\cite{schulman2017proximal}, \emph{Soft actor-critic (SAC)}~\cite{pmlr-v80-haarnoja18b}, etc. We refer readers to~\cite{johnson2020arch} for more details. DRL can be naturally applied to learn control policies for CPS by interacting with its \emph{plant}, which is the physical environment of the system.

\subsection{Temporal Specification of CPS}\label{sec:temporalSpec}

The behavior of a CPS is usually constrained by temporal requirements.  \emph{Signal Temporal Logic (STL)} is a widely adopted specification language to describe such requirements. In the following paragraphs, we briefly explain the syntax and semantics of STL.

\myparagraph{STL Syntax} STL are composed of \emph{atomic propositions} and \emph{formulas}, defined as follows:
\begin{math}
\alpha \,::\equiv\, f(x_1, \dots, x_N) > 0, \;
\varphi \,::\equiv\, \alpha \mid \bot \mid \neg \varphi \mid \varphi_1 \wedge \varphi_2 \mid \varphi_1\, \UntilOp{I} \varphi_2
\end{math}
Here, $f$ is an $N$-ary function $f:\mathbb{R}^N \to \mathbb{R}$, $x_1, \dots, x_N$ are variables,
$I$ is a closed non-singular interval in $\mathbb{R}_{\ge 0}$,
i.e., $I=[a,b]$ or $[a, \infty)$ where $a,b \in \mathbb{R}$ and $a<b$.
%
$\UntilOp{}$ denotes the \emph{until} operator, and 
$\varphi_1\UntilOp{I}\varphi_2$ requires that $\varphi_1$ should be true during the interval $I$ until $\varphi_2$ becomes true.
Other common connectives such as $\vee$, $\rightarrow,\top$, $\Box_{I}$ (always) and $\Diamond_{I}$ (eventually), are introduced as abbreviations: $\varphi_1\vee\varphi_2\equiv \neg\varphi_1\land\varphi_2$, $\varphi_1 \rightarrow \varphi_2 \equiv \neg \varphi_1 \vee \varphi_2$, $\Diamond_{I}\varphi\equiv\top\UntilOp{I}\varphi$ and $\Box_{I}\varphi\equiv\lnot\Diamond_{I}\lnot\varphi$. An atomic formula $f(\vec{x})\le c$, where $c\in\mathbb{R}$, is accommodated using $\lnot$ and the function $f'(\vec{x}):=f(\vec{x})-c$.

\myparagraph{Quantitative Robust Semantics} Traditional temporal logics (\eg, linear temporal logic) only have Boolean semantics that states the Boolean satisfaction of a formula. In contrast, STL is equipped with quantitative semantics~\cite{donze2010robust} that indicates not only \emph{if} a formula is satisfied, but also \emph{how much} a formula is satisfied.

Formally, the STL quantitative semantics $\Robust{s}{\varphi}$ maps a signal $s$ and a STL formula $\varphi$ to a real number, which reflects how robustly $s$ satisfies $\varphi$. The larger this real number is, the further $s$ is from violating $\varphi$. If the number becomes negative, it indicates that $s$ violates $\varphi$. For example, let $s$ be a variable and $\varphi\equiv s>0$ be the specification. Then  $\Robust{s}{\varphi}$ can simply be the value of $s$: once $s$ is negative,  $\varphi$ is violated. If interested, please refer to~\cite{donze2010robust} for the complete definition of STL robust semantics. 

\subsection{CPS Testing Methodology}\label{sec:classicFalsification}
Falsification is an established methodology in CPS testing~\cite{annpureddy2011s, donze2010breach, zhang2018two, dreossi2019compositional}. Given a model $M^C$ and an STL specification $\varphi$, falsification searches for a counterexample input signal $i$ such that the corresponding output signal $M^C(i)$ violates $\varphi$. In this way, it proves the unsatisfiability of the system model $M^C$ to the specification $\varphi$. 

Alg.~\ref{algo:classicFals} describes a basic falsification algorithm. Essentially, this algorithm formulates the testing problem as an optimization one, by taking the robustness as the objective function. In this algorithm, the CPS model $M^C$ is treated as a black box---only its input signal $i$ and the corresponding output signal $M^C(i)$ (system states) can be observed. In the main loop (Lines~\ref{line:BreachIterateI}--\ref{line:BreachBrUpdateRob} in Alg.~\ref{algo:classicFals}), falsification tries different input signals $i_k$ to minimize the robustness value $\rb_k$ computed based on $M^C(i_k)$ and $\varphi$, so that the system is closer to violation of the specification $\varphi$. Once a negative robustness is observed, falsification will terminate and return that input as a counterexamples. Otherwise, it will keep searching until the time budget is run out (Line~\ref{line:BreachBrReturn}).
\begin{figure}
\begin{algorithm}[H]
\caption{The classic falsification algorithm}
\label{algo:classicFals}
\footnotesize
\begin{algorithmic}[1]
\Require CPS model $M^C$, an STL specification $\varphi$, 
a budget $\globalBudget$.
\vspace{0.3em}
\Function{Hill-Climb-Falsify}{$M^C, \varphi, \globalBudget$}
\State initialize a placeholder $i$ \Comment{best input so far} \label{line:BreachBrInitU}
\State $\rb \gets \infty$ \Comment{minimum robustness so far} \label{line:BreachBrInitRob}
\For{$k \in \{0,\ldots, \globalBudget\}$}\label{line:BreachIterateI}
\State $i_k \gets
\Call{Hill-Climb}{\langle i_l, \rb_l\rangle_{l = 0,\ldots, k-1}}\label{line:BreachMutate}
$\label{line:BreachHillClimbing}
\vspace{0.1em}
\State $\rb_k \gets \Robust{M^C(i_k)}{\varphi}$\label{line:BreachComputeRob}
\Comment{compute robustness}
\vspace{0.1em}
\If{$\rb_k < \rb$}
\State $\rb\gets \rb_k$, $i\gets i_k$\label{line:BreachBrUpdateRob}
\EndIf
\EndFor
\State \Return 
$\begin{cases}
i & \text{if } \rb < 0\\
\text{Failure} & \text{otherwise, no violation is found}
\end{cases}
$\label{line:BreachBrReturn}
\EndFunction
\end{algorithmic}
\end{algorithm}
\vspace{-5pt}
\end{figure}

To solve the optimization problem, falsification employs \emph{hill-climbing optimization} algorithms, as shown in Line~\ref{line:BreachHillClimbing} of Alg.~\ref{algo:classicFals}. \emph{Hill-climbing optimization} algorithms are a family of stochastic meta-heuristics-based optimization algorithms. In general, these algorithms first do random sampling in the input space and obtain the robustness values of these inputs. Then based on the observations, they propose new samplings with the aim of decreasing robustness. Typical such algorithms include CMAES~\cite{hansen2003reducing}, Global Nelder-Mead~\cite{luersen2004globalized}, Simulated Annealing~\cite{van1987simulated}, \etc.

\section{Benchmark Collection}\label{sec:benchmarks}
\begin{table*}[!tb]
\centering
\caption{The collected subject CPS}
\label{tab:benchmarks}
\resizebox{\textwidth}{!}{%
\begin{tabular}{llllr}
\toprule
\textbf{Subject CPS} & \textbf{Domain} & \textbf{Description} & \textbf{Type of Controller} & \textbf{\#Blocks} \\\hline
{Adaptive Cruise Control (ACC)} & {Driving assistant} & Maintain a safety distance from a lead car  & {Model predictive control} & {297} \\

Lane Keeping Assistant (LKA) & Driving assistant & Keep a vehicle in the center of a lane   & Model predictive control & {427} \\

Automatic Parking Valet (APV) & Driving assistant & Park a vehicle on a target spot & Model predictive control  &  {224}\\

Exothermic Chemical Reactor (CSTR) & Chemical reactor & Promote a reactor transition of conversion rate  & Model predictive control  & {316} \\

Land a Rocket (LR) & Aerospace  & Land a rocket on a target position  & Model predictive control & {175} \\

Abstract Fuel Control (AFC) & Powertrain & Maintain the reference air-to-fuel ratio  & PI \& Feedforward control & {281} \\

Wind Turbine (WT) & Power grid & Maintain demanded values for key components & PID \& Lookup table &  {161}\\

Steam Condenser (SC) & Thermal systems & Maintain the reference pressure & PID control &  {56}\\

Water Tank (WTK) & Water storage  & Keep the water level at a reference value & PID control & {919} \\

 \bottomrule
\end{tabular}
}
\end{table*}
\subsection{Benchmark Collection} 

As most industrial CPS are treated as private intellectual properties and are kept confidential, it is challenging to find a large collection of practical and open-sourced CPS. We focus on two sources that potentially release open-sourced systems with documentations: (1) the distribution of \matlab control-related toolboxes, such as the \emph{model predictive control} toolbox\footnote{\url{https://www.mathworks.com/products/model-predictive-control.html}}, and (2) CPS-related literature, such as the papers from software engineering and cyber-physical systems.
Eventually, we selected nine industry-level CPS in seven domains based on the following criteria (see Table~\ref{tab:benchmarks}).
\begin{compactitem}[$\bullet$]
    \item {\bf\em Open-source.} A CPS must be open-sourced so that we can modify the system to replace the traditional controller in it with AI controllers.
    \item {\bf\em Documentation.} A CPS must have comprehensive documentation so that we can configure the system properly. Furthermore, to train AI controllers using DRL, we need to understand the system requirements to design proper reward functions in DRL.
    \item {\bf\em Complexity.} A CPS must reflect the industrial complexity in order to get useful insights and implications from the experiments. 
    \item {\bf\em Simulink}. A CPS must use Simulink as its modeling platform. Simulink is a \matlab-based modeling environment developed by \emph{Mathworks} and is widely adopted in industry. Having a consistent platform makes it easier to set up and run the benchmarks.
\end{compactitem}

Table~\ref{tab:benchmarks} gives an overview of the nine CPS in the benchmark. Specifically, Column {\sf \#Blocks} shows the number of blocks in each system, which is often used to measure the complexity of a CPS.
Each system ships with a built-in traditional controller. 
We experiment with different types of learning algorithm, various agent configurations, and diverse reward functions to explore the capability of AI controllers.
Among all variants, we select the best one as the final AI controller for each system. Due to the page limit, we elaborate on three representative systems (of different types) and their traditional controllers and AI controllers in the following section. 

\subsection{Three Representative CPS Examples}

\myparagraph{Adaptive Cruise Control (ACC)} ACC is released by Mathworks~\cite{ACCMathworks}. ACC is deployed in a driving environment with an ego car and a lead car. The goal of this system is to make the ego car move at a user-set velocity \vset as long as the relative distance \drel between two cars is greater than the safety distance \dsafe. The external input of the whole system is the acceleration of the lead car. The outputs include the velocity \vego and the position of the ego car. 



The traditional controller used in this system is \emph{model predictive control (MPC)}. MPC achieves the optimal control at each moment by predicting the motions of the two cars in a finite time-horizon, and optimizing the acceleration of the ego car to maintain the safe distance. Specifically, MPC  uses a linear model to predict the acceleration and velocity of both cars. 

The DRL controller in ACC collects the system environment information to generate observation states. By evaluating the current state and computing a corresponding state value, the DRL controller outputs an acceleration command to the ego car. 
%
Unlike MPC, the DRL controller uses a reward function to evaluate the agent performance from two aspects: velocity and distance. While the safety distance $\dsafe$ is secured, the ego car should approach to the cruise velocity $\vset$.  Otherwise, it follows the lead car velocity to avoid collision. The reward function penalizes the agent by the violation of the safety distance requirement and rewards the agent based on how close the ego car velocity \vego is to the target speed.

\myparagraph{Steam Condenser (SC)}
This system is collected from~\cite{yaghoubi2019gray}. As an 
indispensable part of modern steam power plants, SC is a sealed container where the steam is condensed by cooling water. The goal of the system is to maintain the pressure of condenser $P$ at a desired level \pref. 
The external input of the system is the steam mass flow rate \fs ($kg/s$). The output of the system is the internal pressure of the condenser.
The traditional controller used in SC is a PID controller which outputs a cooling water flowrate. 
A typical PID controller includes three parameters: proportional (P), integral (I) and Derivative (D).  $P$ reflects the current deviation of the system, $I$ mirrors the accumulations of past errors and $D$ represents on the anticipatory control. Through on-demand combination of the above three control parameters, the system error can be corrected.

The DRL controller in SC outputs a cooling water flowrate to maintain the condenser pressure at a desired level. It uses a reward function to make the condenser reach the desired pressure level as soon as possible and maintain this pressure until a new desired pressure is set. 
When the SC system moves to a steady state, the error signal may fluctuate at a small range. Then the reward function amplifies the error to improve the agent performance.

\myparagraph{Abstract Fuel Control (AFC)} AFC is a complex air-fuel control system released by Toyota~\cite{jin2014powertrain}. The whole system takes two input signals from the outside environment, $ \pedal$ and $\engine$, and outputs $\mu = \frac{|\af - \afref|}{\afref}$, which is the deviation of the air-to-fuel ratio \af from a reference value \afref. The goal of this system is to control the deviation $\mu$ no more than a predefined threshold.

The original control system consists of two parts: (1) a PI controller, and (2) a feed-forward controller. The former regulates the air-to-fuel ratio \af in a closed loop, using the measured \af to compute the fuel command. The latter estimates the rate of air flow into the cylinder by measuring of the inlet air mass flow rate. 

The DRL controller in AFC gathers the information about engine dynamics and outputs a fuel command to achieve the reference \af ratio. 
It uses a reward function to guide the agent to reduce the deviation $\mu$. Specifically, a positive reward is given based on how smaller $\mu$ is and a negative feedback is generated if $\mu$ exceeds certain threshold. A small penalty is added based on the DRL action value from the last time step to acquire a stable control output.

\section{Study Design}
As illustrated in Fig.~\ref{figs:workflow}, we perform experiments to answer three research questions. In RQ1, we evaluate the performance of AI-enabled CPS vs.~traditional CPS. In RQ2, we evaluate the effectiveness of the existing falsification methods. In RQ3, we investigate the possibility of combining traditional and AI controllers. All experiments are based on system specifications from the official documentation (\S{}\ref{sec:stlmetrics}). These specifications further derive different metrics and experiment settings for each RQ, which we detail in~\S\ref{sec:RQ}. 



\subsection{System Specifications}\label{sec:stlmetrics}
\begin{table*}[!tb]
\centering
\caption{The STL specifications of the benchmark CPS. 
\spec{1} follows the pattern: $\Box_I(\varphi_1)$; \spec{2} follows the pattern: $\Box_I(\varphi_2)$; \spec{3} follows the pattern: $\Diamond_I\Box_{I'}(\varphi_3)$; \spec{4} follows the pattern: $\Box_I(\varphi_{41}\to\Diamond_{I'}\varphi_{42})$; \spec{5} follows the pattern: $\Diamond_I(\varphi_5)$;
}
\label{tab:evaluationMetrics}
\scriptsize
\setlength{\tabcolsep}{4pt}
\begin{tabular}{cccccc}
\toprule
 Systems & \spec{1}--hard safety & \spec{2}--soft safety &\spec{3}--steady state & \spec{4}--resilience & \spec{5}--liveness  \\\hline
\multirow{3}{*}{\acc} 
& \multirow{3}{*}{\shortstack{$I = [0,50]$\\$\varphi_1 \equiv \drel \ge \dsafe$}} 
& \multirow{3}{*}{\shortstack{$I = [0,50]$\\$\varphi_2 \equiv \vego \le \vset$}} 
& \multirow{3}{*}{\shortstack{$I = [0,50], I'=[0,40]$\\$\varphi_3\equiv \drel \ge \dsafe + 0.2$}} 
& \multirow{3}{*}{\shortstack{$I = [0,50], I' = [0,1]$ \\ $\varphi_{41}\equiv \neg \varphi_3, \varphi_{42} \equiv \varphi_3$}}
& \multirow{3}{*}{\shortstack{$I = [0,50]$\\$\varphi_5 \equiv \vego \ge 1$}} \\ \\ 
 \\ \hline
 
 \multirow{3}{*}{\lka}
 & \multirow{3}{*}{\shortstack{$I = [0,15]$\\$\varphi_1\equiv|\mathit{error_{1}}|\le 0.85$}} 
 & $I = [0,15], \varphi_2\equiv\varphi_{21}\land \varphi_{22}$ 
 & \multirow{3}{*}{\shortstack{$I = [0,15], I'=[0,10]$\\$\varphi_3\equiv|\mathit{error_{1}}|\le 0.5$}} 
& \multirow{3}{*}{\shortstack{$I = [0,15], I' = [0,1]$ \\ $\varphi_{41}\equiv \neg \varphi_3, \varphi_{42} \equiv \varphi_3$}}
 & \multirow{3}{*}{\shortstack{$I = [0,15]$\\$\varphi_5\equiv|\mathit{v}|\ge 1$ }} \\
 &  & $\varphi_{22}\equiv|\mathit{error_{1}}| = 0$ &  &  &  \\
 &  & $\varphi_{21}\equiv|\mathit{error_{2}}| = 0$ &  &  &  \\ \hline

\multirow{3}{*}{\apv} 
& \multirow{3}{*}{\shortstack{$I = [0,12]$\\$\varphi_1\equiv|\mathit{error_{1}}|\le 1$}} 
& $I = [0,12], \varphi_2\equiv\varphi_{21}\land \varphi_{22}$ 
& \multirow{3}{*}{\shortstack{$I = [0,12], I'=[0,10]$\\$\varphi_3 \equiv|\mathit{error_{1}}| \le 0.5$}} 
& \multirow{3}{*}{\shortstack{$I = [0,12], I' = [0,1]$ \\ $\varphi_{41}\equiv \neg \varphi_3, \varphi_{42} \equiv \varphi_3$}}
& \multirow{3}{*}{\shortstack{$I = [0,12]$\\$\varphi_5\equiv|\mathit{v}|\ge 0.1$ }}\\
 &  & $\varphi_{21}\equiv|\mathit{error_{1}}| = 0$  &  &  &  \\
 &  & $\varphi_{22}\equiv|\mathit{error_{2}}| = 0$  &  &  &  \\ \hline
\multirow{3}{*}{\cstr} 
& \multirow{3}{*}{\shortstack{$I = [25,30]$\\$\varphi_1\equiv|\mathit{error}| \le 0.5$}}
& \multirow{3}{*}{\shortstack{$I = [25,30]$\\$\varphi_2\equiv|\mathit{error}| = 0$}} 
& \multirow{3}{*}{\shortstack{$I = [25,30], I' = [0,4]$\\$\varphi_3\equiv|\mathit{error}| \le 0.4$}} 
&\multirow{3}{*}{\shortstack{---}}
&\multirow{3}{*}{\shortstack{---}} \\ 
 &  &  &  &  &   \\
  &  &  &  &  &  \\ \hline
\multirow{3}{*}{\lr} 
& \multirow{3}{*}{\shortstack{$I = [14.8,15]$\\$\varphi_1\equiv|\mathit{error}|\le 0.5$}} 
& \multirow{3}{*}{\shortstack{$I = [14.8,15]$\\$\varphi_2\equiv|\mathit{error}| = 0$}} 
& \multirow{3}{*}{---}
& \multirow{3}{*}{---}
& \multirow{3}{*}{\shortstack{$I = [0,15]$\\$\varphi_5\equiv|\mathit{v}|\ge 0.1$}} \\
 &  &  &  &  &  \\
 &  &  &  &  &  \\ \hline
\multirow{3}{*}{\afc} 
& \multirow{3}{*}{\shortstack{$I=[0,30]$\\$\varphi_1\equiv\mu\le 0.2$}}  
& \multirow{3}{*}{\shortstack{$I=[0,30]$\\$\varphi_2\equiv\mu = 0$}}  
& \multirow{3}{*}{\shortstack{$I=[0,30], I'=[0,20]$\\$\varphi_3\equiv\mu\le 0.1$}} 
& \multirow{3}{*}{\shortstack{$I = [0,30], I' = [0,1]$ \\ $\varphi_{41}\equiv \neg \varphi_3, \varphi_{42} \equiv \varphi_3$}}  & \multirow{3}{*}{---} \\
 &  &  &  &  &  \\
 &  &  &  & & \\  \hline
\multirow{3}{*}{\wt} 
& \multirow{3}{*}{\shortstack{$I=[30,630]$\\$\varphi_1\equiv\theta\le 14.2$}}  
& \multirow{3}{*}{\shortstack{$I=[30,630]$\\$\varphi_2\equiv\theta\ = 0$ }}  
& \multirow{3}{*}{\shortstack{$I = [30,630], I' = [0,500]$ \\$\varphi_3\equiv\theta\le 14$}}
& \multirow{3}{*}{\shortstack{$I = [30,630], I' = [0,5]$ \\ $\varphi_{41}\equiv \neg \varphi_3, \varphi_{42} \equiv \varphi_3$}}
& \multirow{3}{*}{---} \\
 &  &  &  &  &  \\
 &  &  &  &  &  \\  \hline
\multirow{3}{*}{\scd} 
& \multirow{3}{*}{\shortstack{$I=[30,35]$\\$\varphi_1 \equiv |\mathit{error}| \le 0.5$}}  
& \multirow{3}{*}{\shortstack{$I=[30,35]$\\$\varphi_2\equiv |\mathit{error}| = 0$}}  
& \multirow{3}{*}{\shortstack{$I = [30,35], I' = [0,4]$\\$\varphi_3\equiv|\mathit{error}| \le 0.4$}} 
& \multirow{3}{*}{---}
& \multirow{3}{*}{---} \\
 &  &  &  &  &  \\
 &  &  &  &  &  \\  \hline
\multirow{3}{*}{\wtk}
& \multirow{3}{*}{\shortstack{$I=[5,6] \cup [11,12] \cup [17, 18]$\\$\varphi_1\equiv\ |\mathit{error}|  \le 0.2$ }} 
& \multirow{3}{*}{\shortstack{$I=[5,6] \cup [11,12] \cup [17, 18] $\\$\varphi_2\equiv\ |\mathit{error}| \ = 0$ }} 
& $I=[5,6] \cup [11,12] \cup [17, 18]$
& \multirow{3}{*}{---}
& \multirow{3}{*}{---}\\
 &  &  & $ I' = [0,0.8]$  &  &  \\
 &  &  & $\varphi_3\equiv|\mathit{error}| \le 0.15$ &  & \\
 \bottomrule
\end{tabular}
\end{table*}
In this work, we adopt STL (introduced in~\S{}\ref{sec:temporalSpec}) as our specification language to evaluate the temporal properties of CPS. Specifically, we extract the temporal properties of the models from their documents, and summarize them in Table~\ref{tab:evaluationMetrics}. 
 We classify these properties into 5 categories, namely \boldspec{1}-\boldspec{5}, according to their semantics:

\begin{compactitem}[$\bullet$]

\item \textbf{\textit{\boldspec{1}: Hard safety}.} \spec{1} must be strictly satisfied by the systems, any violation of \spec{1} can lead to safety problems. \spec{1} follows the pattern $\Box_I(\varphi_1)$, where $\varphi_1$ is a system invariant during the simulation. 

\item \textbf{\textit{\boldspec{2}: Soft safety}.} \spec{2} follows the similar pattern $\Box_I(\varphi_2)$ as \spec{1}, but the satisfaction of $\varphi_2$ is not demanded. Instead, \spec{2} is used to measure the average and maximum deviations of the outputs from the reference values. Based on that, we can understand the average and boundary behaviors of the controllers. 

\item \textbf{\textit{\boldspec{3}: Steady state}.} \spec{3} follows the pattern $\Diamond_I\Box_{I'}(\varphi_3)$. It requires the system to satisfy $\varphi_3$ for the interval $I'$ at some point in $I$. \spec{3} is used to evaluate if the system reaches the steady state $\varphi_3$ and stays there for the interval $I'$. 

\item \textbf{\textit{\boldspec{4}: Resilience}.} \spec{4} follows the pattern $\Box_I(\varphi_{41}\to\Diamond_{I'}\varphi_{42})$. It requires that, during the interval $I$, whenever an event $\varphi_{41}$ happens, the system should react by satisfying $\varphi_{42}$ within $I'$. \spec{4} is used to evaluate the system's ability to recover from fluctuations. 

\item \textbf{\textit{\boldspec{5}: Liveness}.} \spec{5} follows the pattern $\Diamond_I(\varphi_5)$, where $\varphi_5$ should be eventually satisfied during $I$. \spec{5} is used to inspect the system to avoid the case when the controller makes conservative control decisions to passively meet the safety requirements. 

\end{compactitem}


\subsection{Research Questions}\label{sec:RQ}

\vspace{3pt}
\noindent\textbf{RQ1. How well do the DRL-based AI controllers perform compared with the traditional controllers?} 
\vspace{3pt}

 
While many AI controllers have been proposed and used, there has not been a systematic study on how well AI controllers perform compared with traditional controllers. RQ1 aims to compare the performance of these two kinds of controllers and understand their \emph{pros and cons}. Since DRL-based approaches are the state of the art among current AI controllers, we mainly focus on DRL-based controllers in this study. 
In the experiment, we first randomly generate 100 input signals. Then we run simulations on each of the CPS with the traditional controller and the DRL-based controller respectively. We compare the performances of these controllers according to multiple properties introduced in~\S{}\ref{sec:stlmetrics}. To better understand the quality of the controllers, we not only consider Boolean satisfactions to the STL formulas in~\S{}\ref{sec:stlmetrics}, but also propose a series of more fine-grained metrics that consider the semantics of different formula patterns. These fine-grained metrics for \boldspec{1}-\boldspec{5} are listed as follows:

\begin{compactitem}[$\bullet$]

\item \textbf{\boldspec{1}}: We record the number of satisfying simulations $n_s$ and compute the satisfaction ratio over 100 rounds of simulation since \spec{1} are safety properties required to be satisfied strictly. 

\item \textbf{\boldspec{2}}: In addition to the Boolean satisfaction of \spec{2}, we are also interested in how much \spec{2} is violated in each case. Hence, given an output signal $s$, we first collect all the moments $T$
when $\varphi_2$ is violated. Then, we present two metrics regarding \spec{2}:
1) \textbf{MAE}---the mean  value of the average absolute error over 100 simulations, and
2) \textbf{MAXERR}---the mean value of the maximum absolute error over 100 simulations.
With these two metrics, we can understand the average and the extreme violations, respectively.

\item \textbf{\boldspec{3}}: We use \spec{3} to extensively explore how stable the system is during each round of simulation. For 100 simulations, we record how many moments $T$ are there when $s(t)$ falls into a steady state defined by $\varphi_3$. Then we compute the percentage of these steady state moments by $\tfrac{\#T}{\#T_{\mathit{total}}}$ in each round of simulation. We take the average value over 100 simulations as the fine-grained metric to evaluate \spec{3}. 

\item \textbf{\boldspec{4}}: We use $\frac{\nr}{\nv}$ to measure how good the systems satisfy \spec{4} during one round of simulation, where $\nv$ is the number of cases when the system falls into the fluctuating state defined by $\varphi_{41}$, and $\nr$ is the number of the cases when the system can react by doing $\varphi_{42}$. We then take the average value of $\frac{\nr}{\nv}$ over 100 simulations.

\item \textbf{\boldspec{5}}: We record the number of simulation rounds $n_s$  satisfying \spec{5} in~\S{}\ref{sec:stlmetrics}. Then we calculate the ratio of satisfaction over 100 rounds of simulation.

\end{compactitem}
By analyzing the evaluation results with the metrics above, we can obtain more comprehensive information to understand the advantage and limitations of DRL-based AI controllers and traditional controllers, from multiple perspectives.

\vspace{5pt}
\noindent\textbf{RQ2. To what extent are existing CPS testing methods still effective on AI-based CPS?}
\vspace{3pt}

For RQ2, we focus on an established CPS testing methodology called falsification (introduced in \S{}~\ref{sec:classicFalsification}).  Falsification has proved to be effective on traditional CPS~\cite{yamagata2020falsification, zhang2018two, adimoolam2017classification}, but few studies have evaluated it on AI controllers. Since falsification methods are guided by logic semantics, it is unclear whether it is still effective on AI controllers, which are essentially statistical methods with high uncertainty and low interpretability.

In the experiment, we select two widely-used falsification tools, Breach~\cite{donze2010breach} and S-TaLiRo\cite{annpureddy2011s}. 
Note that both Breach and S-TaLiRo 
integrate several different back-end optimization solvers, and apply different tricks to improve the performance. We select Global Nelder-Mead (GNM) and CMAES for Breach, and Simulated Annealing (SA) and stochastic optimization with adaptive restart (SOAR)~\cite{mathesen2019falsification} for S-TaLiRo according to the findings in~\cite{ernst2020arch}.

Due to the stochasticity of the falsification algorithms, for each experiment, we repeatedly run the falsification algorithms, and obtain a falsification rate $\frac{\text{\#~successful trials}}{\text{\#~total trials}}$ as the evaluation metric. We also report the average time consumption, and the average number of successful falsification trials as complementary metrics.



In this experiment, we only run falsification on the systems that never violate \spec{1} (hard safety) during random simulations in the experiment of RQ1. Because these systems are hard to be evaluated in RQ1, which can be used as the changeable samples to measure the performance difference among different falsification approaches.

Our results of RQ1 and RQ2 form a basic benchmark quality and reliability analysis of AI-enabled CPS.

\vspace{5pt}
\noindent\textbf{RQ3. Can the combination of traditional and AI controllers bring better performance?}
\vspace{3pt}

As suggested by international standards \emph{ISO26262}~\cite{iso26262} and \emph{ISO/PAS 21448 (SOTIF)}~\cite{iso21448}, 
 modular redundancy (\eg, doubling or tripling) is an important way to improve system quality and reliability.
Therefore, in RQ3, we aim to investigate the performance of hybrid controllers---a novel type of controllers which combines the traditional and DRL-based ones. 
We perform an exploration on three typical approaches for such combinations: (1) a random-based approach, (2) an average-based approach, and (3) a scenario-dependent approach, with the purpose to understand whether this could be a promising direction for further research. 
\begin{compactitem}[$\bullet$]

\item \textbf{\textit{Random-based}.} This method chooses a signal randomly from two controllers and passes it to the subsequent components. Here, \emph{sample time} is a hyperparameter that indicates how frequently the controller switches. We use 2 different sample times in our experiments: (1) 0.1 sec, which is the minimum system step time, and (2) 1 sec, which is a log scale increase for comparison.

\item \textbf{\textit{Average-based}.} This method takes the average value of the outputs of two controllers as the final output.

\item \textbf{\textit{Scenario-dependent}.} 
Based on the insights from the experiments of RQ1 and RQ2, we summarize the scenarios where different types of controllers perform well and design a dynamic controller switch logic to achieve the optimal control strategy.
For instance, for a system with controller A and B, A may have smaller value on S2 (averaged error signal) while B may have better performance on S4 (resilience). A switch logic can be: if the error signal exceeds a specified threshold which indicates the system falls into an unsteady state, the control will be granted to B to quickly recover to the steady state, and A may take in charge to maintain a small error value under the steady state.
\end{compactitem}

In RQ3, we use the same metrics from RQ1 and RQ2 to evaluate these hybrid controllers under the same experimental settings.

\myparagraph{Hardware \& Software Dependencies} DRL training is computationally intensive, so we use a server with 3.5GHz Intel i9-10920X CPU, 15GB RAM, and an NVIDIA TITAN V GPU. Other experiments (\eg, falsification) were conducted using Breach 1.9.0 with GNM and CMAES as solvers and S-TaLiRo 1.6.0 with SA and SOAR as solvers on an Amazon EC2 c4.xlarge server with 2.9GHz Intel Xeon E5-2666 CPU, 4 virtual CPU cores, and 8GB RAM.

\section{Experimental Results}

\subsection{RQ1. Performance of AI-enabled CPS vs. ~traditional CPS}
\begin{figure*}[!tb]
    \centering
    \includegraphics[width=\linewidth]{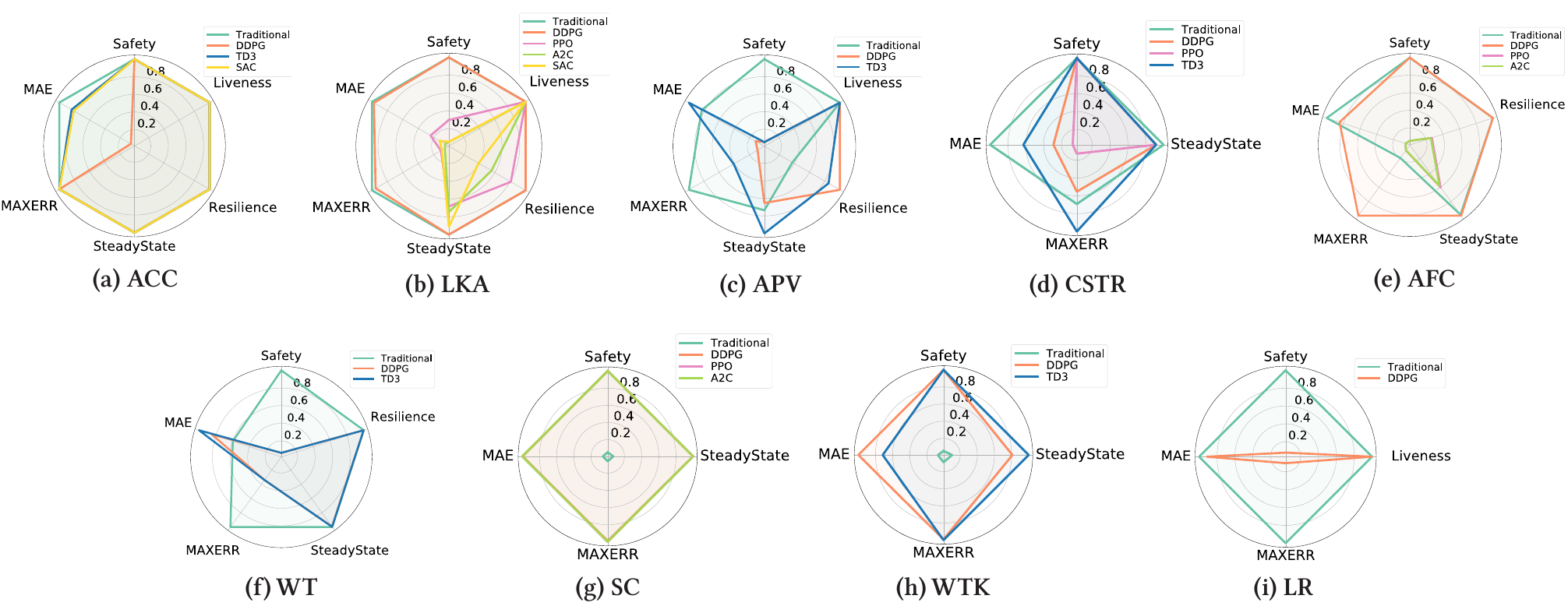}
    \caption{Performance comparison on systems with traditional and DRL controllers (RQ1)}
    \label{fig:rq1}
\end{figure*}
Fig.~\ref{fig:rq1} summarizes the evaluation results of  traditional and AI-enabled CPS via radar charts. Each axis in a radar chart represents an evaluation metric from~\S{}\ref{sec:stlmetrics}. We normalize the result of each metric into $[0,1]$. A higher value indicates a better performance.
\begin{compactitem}[$\bullet$]
\item \textit{\spec{1}: Hard safety}. 
Not surprisingly, traditional controllers have better or at least comparable performance on \spec{1} in 7 of the 9 systems ($\acc$, $\lka$, $\apv$, $\cstr$, $\lr$, $\afc$, and $\wt$), compared to DRL controllers. This is because safety  always has the highest priority in the design of a traditional controller.
However, in \wtk and \scd, the traditional PID controllers fail to regulate the error signals within a limited range. In particular, an oscillated signal can be generated by PID controllers and the maximum error threshold is therefore violated by an instantaneous overshoot. 

DRL controllers also have good performance on \spec{1} in 6 systems ($\acc$, $\lka$, $\cstr$, $\afc$, $\scd$, and $\wtk$). However, DRL controllers are better than traditional controllers in $\scd$ and $\wtk$, while worse in $\wt$ and $\apv$.  Unlike the behavior of PID controllers in $\scd$ and $\wtk$, DRL controllers output stable control signals with no instant overshoot. Thus, the safety requirements are held.

Overall, for \spec{1}, traditional controllers and DRL controllers have similar performance in systems like $\acc$, $\lka$, $\cstr$, and $\afc$. Also, for all systems, there is always at least one controller that never violates \spec{1}. These results indicate that, the random sampling method cannot effectively evaluate systems on \spec{1}, and therefore, we use the more advanced testing method, namely falsification, in RQ2 to acquire more in-depth insights.
\item \textit{\spec{2}: Soft safety}.
The traditional controllers have good MAE performance in 5 of the 9 systems ($\acc$, $\lka$, $\cstr$,$\lr$, and $\afc$) and decent MAXERR performance in 5 systems ($\acc$, $\lka$, $\apv$,$\lr$, and $\wt$). 
Compared with the results on \spec{1}, there are fewer traditional CPS that outperform AI-enabled CPS on \spec{2}. Moreover, a traditional controller with a good MAE result may not have a similarly good MAXERR result, \eg, \afc and \cstr. This indicates that the output signals from traditional controllers could be unstable, compared with DRL controllers. 

DRL controllers bring good MAE results in 7 systems ($\acc$, $\lka$, $\apv$, $\wt$, $\scd$, $\lr$, and $\wtk$) and good MAXERR results in 6 systems ($\acc$, $\lka$, $\cstr$, $\afc$, $\scd$, and $\wtk$). 
According to these results, 
if a system has too many performance requirements or multiple control outputs,
such as $\apv$ and $\wt$, a standalone DRL controller may not handle these situations well.

\item \textit{\spec{3}: Steady state}.
Most traditional controllers can provide relatively stable outputs in all systems except $\scd$ and $\wtk$. 
The poor performances of these two systems can be attributed to their oscillated PID outputs. 
In contrast, DRL controllers show good performance in all of the systems. Moreover, we find that different DRL controllers of the same system may have a variance in their performance. For instance, in $\wtk$, while \tdthr does not perform as well as \ddpg in MAE, it outperforms \ddpg in \spec{3}. 
\item \textit{\spec{4}: Resilience}.
Both DRL and traditional controllers show good performances regarding resilience in the 5 applicable systems, except the traditional and \tdthr controllers in $\apv$. This reveals their weaknesses in handling fluctuations. In contrast, although \ddpg performs strongly in resilience, it does not conform to \spec{1}, the hard safety requirements. For that reason, it can not  be considered as a reliable system.
\item \textit{\spec{5}: Liveness}.
All the controllers show positive results on \spec{5}. This indicates that none of our controllers takes conservative strategies to passively satisfy the safety requirements. 
\end{compactitem}
Regarding the comparison between the traditional controller and DRL controllers in each system, DRL controllers have better or similar performance in most of the systems.  However, this is not always the case. For example, in $\apv$ and $\wt$, the DRL-based controllers do not perform as well as their traditional counterparts. 


\begin{tcolorbox}[size=title, colback=white]
{\textbf{Answer to RQ1:} Though AI controllers achieve better or comparable performance in many cases, they also fail to properly function in several complex cases, implying that falling back to traditional controllers can be a safer choice in some complex scenarios that involve multiple control requirements.}
\end{tcolorbox}

\begin{table*}[!tb]
\caption{Falsification performance comparison between four existing falsification algorithms (RQ2)}
\vspace{-10pt}
\label{tab:rq2}
\begin{subtable}[t]{\textwidth}
\resizebox{\textwidth}{!}{%
\begin{tabular}{crrr|rrr|rrr|rrr|rrr|rrr|rrr|rrr}
\hline
         & \multicolumn{3}{c}{\acc-\trad} & \multicolumn{3}{c}{\acc-\ddpg} & \multicolumn{3}{c}{\acc-\tdthr} & \multicolumn{3}{c}{\acc-\sac} & \multicolumn{3}{c}{\cstr-\trad} & \multicolumn{3}{c}{\cstr-\ddpg} & \multicolumn{3}{c}{\cstr-\ppo} & \multicolumn{3}{c}{\cstr-\tdthr} \\
         &    FR     &    time    &    \#sim    &    FR    &   time     &   \#sim     &   FR  & time  &   \#sim    &   FR   &  time     &   \#sim    &   FR   &    time   &   \#sim    &  FR    &   time    &  \#sim     &  FR    &    time   &  \#sim   &    FR     &    time    &    \#sim          \\\hline
GNM-BR  
& 3 & 269.9 & 187.7 
& 1 & 250.6 & 144.0 
& \cellcolor[gray]{0.8} 25 &  \cellcolor[gray]{0.8} 162.6 & \cellcolor[gray]{0.8} 107.9 
& 1 & 313.8 & 159.0
& 28 & 292.0 & 56.7
& 0  &  -    &-  
& 0  &  -    &-  
& 0  &  -    &-   \\
CMAES-BR
&  \cellcolor[gray]{0.8} 16 & \cellcolor[gray]{0.8} 142.4 & \cellcolor[gray]{0.8} 101.5 
& 0	& - & -     
& 16 & 76.9 & 30.8  
& \cellcolor[gray]{0.8}  9 & \cellcolor[gray]{0.8}  263.8&  \cellcolor[gray]{0.8} 98.9 
& \cellcolor[gray]{0.8}  30 &  \cellcolor[gray]{0.8}190.8 &  \cellcolor[gray]{0.8} 35.8
& 0	& - & -    
& 0	& - & -    
& 0	& - & -   \\
SA-ST   
& 2 & 148.1 & 148.0
&  \cellcolor[gray]{0.8} 1 &  \cellcolor[gray]{0.8}  218.6 &  \cellcolor[gray]{0.8} 147.0    
& 0 & -     & -
& 1	& 433.7 &	272.0
& 30 & 200.29 & 47.57
& 0	& - & -  
& 0	& - & -  
& 0	& - & -  \\
SOAR-ST
&   0     & -   &  -
&   0     & -  &  -
&   0    & -  &  - 
&   0    & -  &  - 
& 30 & 	214.0& 	50.5	 
&   0    & -      & - 
&   0    & -      & - 
&   0    & -      & -   \\
\hline
\end{tabular}
}
\end{subtable}

\begin{subtable}[t]{\textwidth}
\resizebox{\textwidth}{!}{%
\begin{tabular}{crrr|rrr|rrr|rrr|rrr|rrr|rrr|rrr}
\hline
         & \multicolumn{3}{c}{\afc-\trad} & \multicolumn{3}{c}{\afc-\ddpg} & \multicolumn{3}{c}{\afc-\ppo} & \multicolumn{3}{c}{\afc-\atwoc} & \multicolumn{3}{c}{\scd-\trad} & \multicolumn{3}{c}{\scd-\ddpg} & \multicolumn{3}{c}{\scd-\ppo} & \multicolumn{3}{c}{\scd-\atwoc} \\
         &    FR     &    time    &    \#sim    &    FR    &   time     &   \#sim     &   FR  & time  &   \#sim    &   FR   &  time     &   \#sim    &   FR   &    time   &   \#sim    &  FR    &   time    &  \#sim     &  FR    &    time   &  \#sim   &    FR     &    time    &    \#sim          \\\hline
GNM-BR  
& 0	& -  & -
& 0 & -  & -
& \cellcolor[gray]{0.8}  30& \cellcolor[gray]{0.8} 9.5&  \cellcolor[gray]{0.8} 6.8
& \cellcolor[gray]{0.8}  30& \cellcolor[gray]{0.8}  9.4&  \cellcolor[gray]{0.8} 6.1
& 30& 0.3& 1.0
&  0& - & - 
&  0& - & - 
&  0& - & -      \\
CMAES-BR  
& 0 & - & -
& 0	& - & -
& 10 & 37.1 & 28.9
& 4	 & 77.3 & 61.3
& 30& 0.3& 1.0
&  0& - & - 
&  0& - & - 
&  0& - & -    \\
SA-ST   
&9	& 295.2 & 148.4
& 0	& -	& -
& 0	& -	& - 
& 0	& -	& -
&30	&0.2&	1.0
& 0	& -	& -
& 0	& -	& - 
& 0	& -	& -  \\
SOAR-ST 
& \cellcolor[gray]{0.8} 11 & \cellcolor[gray]{0.8} 411.3  & \cellcolor[gray]{0.8} 157.4
&0	& -	& - 
&0  & - & -
&0  &-  &- 
&  \cellcolor[gray]{0.8} 30	& \cellcolor[gray]{0.8} 0.2& \cellcolor[gray]{0.8} 1.0
& 0	& -	& -
& 0	& -	& -
& 0	& -	& - \\
\hline
\end{tabular}
}
\end{subtable}

\begin{subtable}[t]{\textwidth}
\resizebox{\textwidth}{!}{%
\begin{tabular}{crrr|rrr|rrr|rrr|rrr|rrr|rrr|rrr}
\hline
         & \multicolumn{3}{c}{\lka-\trad} & \multicolumn{3}{c}{\lka-\ddpg} & \multicolumn{3}{c}{\lka-\ppo} & \multicolumn{3}{c}{\lka-\atwoc} & \multicolumn{3}{c}{\lka-\sac} & \multicolumn{3}{c}{\apv-\trad} & \multicolumn{3}{c}{\apv-\ddpg} & \multicolumn{3}{c}{\apv-\tdthr} \\
         &    FR     &    time    &    \#sim    &    FR    &   time     &   \#sim     &   FR  & time  &   \#sim    &   FR   &  time     &   \#sim    &   FR   &    time   &   \#sim    &  FR    &   time    &  \#sim     &  FR    &    time   &  \#sim   &    FR     &    time    &    \#sim          \\\hline
GNM-BR  
& 30 & 636.9 & 75.4
& 0	 & - & -
& 30 & 10.6	& 10.9
& 30 & 5.0  & 5.0
& 30 & 60.5 & 54.0
& 0     & -     & -  
& \cellcolor[gray]{0.8} 30	&\cellcolor[gray]{0.8} 41.5	& \cellcolor[gray]{0.8} 1.0   
& \cellcolor[gray]{0.8} 30	& \cellcolor[gray]{0.8} 42.8   &\cellcolor[gray]{0.8} 1.0  \\
CMAES-BR
&  27	&  414.9&  47.8
&  0	&   -	&  -
&  28	&  21.6	&  16.6
&  30	&  9.4	&  7.5
&  28	&  82.3	&  44.4
& 0      & -      & -  
& 30	&51.6	& 1.0   
& 30	&59.7   & 1.0 \\
SA-ST  
&  24	&1006.2	 &114.8
&  0	&  -  &   - 
&  29	&  28.4	&  33.0
&  29	&  62.8	&  64.0
& \cellcolor[gray]{0.8}  30  & \cellcolor[gray]{0.8}  20.7	& \cellcolor[gray]{0.8}  25.1  
& 0     & -     & -  
& 30    & 112.5 & 1.0
& 30    & 119.2 & 1.0  \\
SOAR-ST  
&  \cellcolor[gray]{0.8} 30   & \cellcolor[gray]{0.8} 262.4   &  \cellcolor[gray]{0.8} 30.8	
&  0     & -       &-
& \cellcolor[gray]{0.8} 30    & \cellcolor[gray]{0.8} 5.2  &  \cellcolor[gray]{0.8} 7.1
& \cellcolor[gray]{0.8}30	    & \cellcolor[gray]{0.8}3.3	& \cellcolor[gray]{0.8} 5.2
&30	&	50.2	&27.6	
& 0      & -      & -  
& 30    & 115.9  & 1.0
& 30    & 126.2   & 1.0 \\
\hline
\end{tabular}
}
\end{subtable}

\begin{subtable}[t]{\textwidth}
\resizebox{\textwidth}{!}{%
\begin{tabular}{crrr|rrr|rrr|rrr|rrr|rrr|rrr|rrr}
\hline
         & \multicolumn{3}{c}{\wt-\trad} & \multicolumn{3}{c}{\wt-\ddpg} & \multicolumn{3}{c}{\wt-\ppo} & \multicolumn{3}{c}{\wtk-\trad} & \multicolumn{3}{c}{\wtk-\ddpg} & \multicolumn{3}{c}{\wtk-\tdthr} & \multicolumn{3}{c}{\lr-\trad} & \multicolumn{3}{c}{\lr-\ddpg} \\
         &    FR     &    time    &    \#sim    &    FR    &   time     &   \#sim     &   FR  & time  &   \#sim    &   FR   &  time     &   \#sim    &   FR   &    time   &   \#sim    &  FR    &   time    &  \#sim     &  FR    &    time   &  \#sim   &    FR     &    time    &    \#sim          \\\hline
GNM-BR   
& \cellcolor[gray]{0.8} 30 & \cellcolor[gray]{0.8} 197.8 & \cellcolor[gray]{0.8} 42.9
& 0 & - & -
& 0 & - & -
&  \cellcolor[gray]{0.8} 30& \cellcolor[gray]{0.8}  3.3 & \cellcolor[gray]{0.8}  5.8
&  0 &  -	&  -
&  0 &  -	&  -
&\cellcolor[gray]{0.8}  20 & \cellcolor[gray]{0.8} 4224.7 &	\cellcolor[gray]{0.8} 80.8
& 30 & 68.5   &	1.0     \\
CMAES-BR 
& 0 & - & -
& 0 & - & -
& 0 & - & -
& 30 & 9.8 & 17.6
&  0 &  -  &  -
&  0 &  -  &  -
&  16	&  4341.1	&  78.4
&  \cellcolor[gray]{0.8} 30	& \cellcolor[gray]{0.8}  54.9	 & \cellcolor[gray]{0.8}  1.0    \\
SA-ST  
& 20& 	335.5& 	65.4
&  0      & -       &-  
&  0     & -      & -
& 30	&  8.8	& 20.6
&  0    &  -  &  -
&  0    &  -  &  -
&13  &6532.2   & 92.9 
&  30   &78.4   & 1.0    \\
SOAR-ST 
&  30& 255.7 & 	50.9   
&  0 &  -  &  -
&  0 &  -  &  -
& 30	&4.0	&9.1
&  0     & -      & -  
&  0     &  -     &  -

& 9   &  7809.1  &    93.2
& 30   &  83.7   &   1.0    \\
\hline
\end{tabular}
}
\end{subtable}
\end{table*}
\subsection{RQ2. Effectiveness of Falsification}

Table~\ref{tab:rq2} shows the experimental results of RQ2, where we evaluate the performances of 4 falsification approaches using the metrics of FR (/30), time (secs), and \#sim (see ~\S{}\ref{sec:RQ}). For each benchmark, we select their \spec{1}, namely hard safety, as the target system specification. The reason is that, according to RQ1, we know that most of the systems do not violate their hard safety properties. Therefore, it makes sense to further investigate the satisfaction of each system to \spec{1}. In our experiments, we set the budget ($\globalBudget$ in Alg.~\ref{algo:classicFals}) as 300. We highlight the best performer for each CPS model in Table~\ref{tab:rq2}, according to their FR.

The results presented in Table~\ref{tab:rq2} reveal apparent differences in the abilities of different falsification algorithms for specific cases. For example, CMAES-BR performs well in \acc-\trad, but does not perform well in \wt-\trad or \afc-\atwoc; SOAR-ST performs well in \afc-\trad, but does not perform well in \acc-\tdthr or \afc-\atwoc. 


We identify a model as \emph{falsifiable} once there exists a falsification algorithm that manages to falsify the model. Based on our observations, we find that, for any falsification algorithm $\mathtt{A}$, there always exists a falsifiable benchmark that can not be falsified by $\mathtt{A}$. For instance, GNM-BR cannot falsify the falsifiable model \afc-\trad, and performs poorly on \acc-\sac; CMAES-BR cannot falsify the falsifiable model \acc-\ddpg or \wt-\trad; SA-ST cannot falsify the falsifiable model \afc-\ppo or \afc-\atwoc; SOAR-ST cannot falsify the falsifiable model \afc-\ppo or \afc-\atwoc. This result proves the famous \emph{no free lunch theorem} in optimization~\cite{wolpert1997no}, and also motivates us to develop more effective algorithms to handle these emerging AI-enabled CPS.


\begin{tcolorbox}[size=title, colback=white]
{\textbf{Answer to RQ2:} Falsification, an established testing method, fails to handle several AI-enabled CPS, highlighting the opportunity to develop new AI-aware testing methods for CPS.}
\end{tcolorbox}

\subsection{RQ3. Performance of Hybrid Controllers}
\begin{figure*}
\centering
\includegraphics[width=0.9\linewidth]{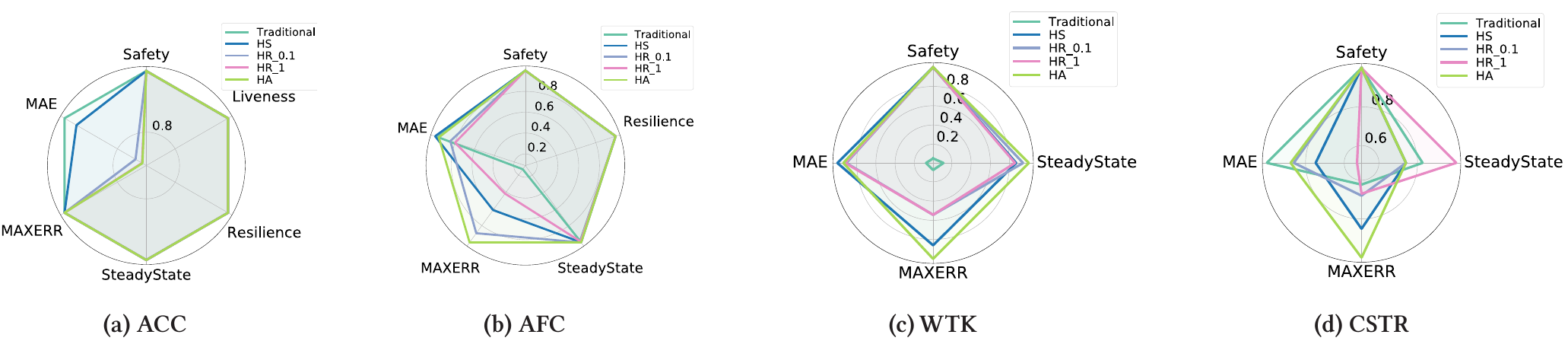}
\vspace{-10pt}
\caption{Performance comparison on systems with hybrid controllers (RQ3)}
\label{fig:rq3}
\end{figure*}
\begin{table}[!tb]
\caption{Performance comparison on 4 hybrid benchmark CPS with falsification tools (RQ3)}
\vspace{-5pt}
\label{tab:rq3}
  \scalebox{0.64}{
\begin{tabular}{c|rrr|rrr|rrr|rrr}
\hline
Tool  & \multicolumn{3}{c|}{GNM-BR} & \multicolumn{3}{c|}{CMAES-BR}  & \multicolumn{3}{c|}{SA-ST}  & \multicolumn{3}{c}{SOAR-ST} \\ \hline
Benchmark  & FR & time & \multicolumn{1}{c|}{\#sim} & FR & time & \multicolumn{1}{c|}{\#sim} & FR                   & time                 & \multicolumn{1}{c|}{\#sim} & FR     & time     & \#sim    \\ \hline
\acc-\hb
&0	    &-      &-	
&11	    &177.9  &71.5
&2  	&465.4  &188.0
&0	    &-	    &-      \\
\acc-\hbra  
&1		&123.9	&48.0
&0		&-  	&-
&1		&392.1	&183.0	
& 0      & -      & -       \\
\acc-\hbrb    
&3	    &313.1  &121.0
&1	    &66.2	&26.0
&1	    &457.6  &209.0								
&0      &-      &-            \\
\acc-\hba 
&4		&437.1	&165.0
&1		&178.9	&72.0
&1		&372.3	&149.0
&0		&-      &-		 \\ \hline
\afc-\hb     
&0		&-  	&-
&0		&-  	&-
&0		&-  	&-
&0		&-  	&-      \\
\afc-\hbra  
&0		&-  	&-
&0		&-  	&-
&0		&-  	&-
&0		&-  	&-       \\
\afc-\hbrb    
& 29	& 148.5	& 87.4
& 30	& 105.1	& 61.0
&0		&-  	&-
&0		&-  	&-		    \\
\afc-\hba 
&0		&-  	&-
&0		&-  	&-
&0		&-  	&-
&0		&-  	&-      \\ \hline
\wtk-\hb       
&0		&-  	&-
&0		&-  	&-
&0		&-  	&-
&0		&-  	&-         \\
\wtk-\hbra  
&0		&-  	&-
&0		&-  	&-
&0		&-  	&-
&0		&-  	&-          \\
\wtk-\hbrb       
&0		&-  	&-
&0		&-  	&-
&0		&-  	&-
&0		&-  	&-        \\
\wtk-\hba 
&0		&-  	&-
&0		&-  	&-
&0		&-  	&-
&0		&-  	&-           \\ \hline
\cstr-\hb  
&3		&835.8	&155.0
&1		&1310.6 &292.0	
&10	    &997.7  &213.4
&11	    &973.5  &172.1		      \\
\cstr-\hbra  
&0		&-  	&-
&0		&-  	&-
&0		&-  	&-
&0		&-  	&-          \\
\cstr-\hbrb    
&22		&411.2	&87.6
&20		&155.8	&33.2
&18		&578.7	&124.0
&30		&369.0	&61.9	      \\
\cstr-\hba 
&0      &-      &-
&0		&-  	&-
&0		&-  	&-	
&1	    &1940.4	&189.0       \\ \hline
\end{tabular}
}
\vspace{-5pt}
\end{table}
%
From the results of RQ1 and RQ2, we find that 
traditional and AI controllers have their own advantages in satisfying different requirements.
For example, in \cstr (Fig.~3d), although the traditional controller performs well in MAE, it does not perform well in MAXERR. In contrast, the \tdthr controller complements the situation, holding a high MAXERR but low MAE. This inspires us to combine traditional and AI controllers to obtain hybrid controllers, which may outperform each of them alone. We explore three combination methods, namely the random-based, the average-based, and the scenario-dependent ones, introduced in~\S{}\ref{sec:RQ}. For the random-based methods, we evaluate on multiple instances varied by their sampling time. Our results include Fig.~\ref{fig:rq3} that uses the same evaluation metrics as RQ1, and Table~\ref{tab:rq3} that applies the 4 falsification methods to those hybrid controllers. 


\begin{compactitem}[$\bullet$]
\item \textit{\acc}.
%
From the Fig.~\ref{fig:rq3} and Table~\ref{tab:rq3}, in $\acc$, all the hybrid controllers perform better in falsification than the traditional controllers, and the scenario-dependent hybrid controller performs well in MAE compared to the traditional controller. 
In contrast, the performances of the random-based or the average-based hybrid controllers are not as good as the scenario-dependent ones.
\item \textit{\afc}.
Among the 4 types of hybrid controllers we deployed, 3 of them have similar or better performance than the traditional one. 
\item \textit{\wtk}.
The scenario-dependent and the average-based hybrid controllers are significantly better than the random-based controllers, and none of the hybrid controllers have been falsified. 
\item \textit{\cstr}.
We find that, the random-based method controller HR\_0.1 and the average-based controller are not falsified. Moreover, the average-based controller also performs well in MAXERR as its constituent $\tdthr$ controller does. 
\end{compactitem}
In summary, the hybrid controllers can take advantage of their constituent controllers.
The results reflect that the random-based combination method with large sample time can lead to an inconsistent control logic, and it can cause serious safety issues. The average-based method can be applied if the system is too complicated to design a controller switching logic and/or the constituent controllers have distinct advantages in different aspects. The scenario-dependent approach is recommended as it can switch between different controllers based on  real-time system states. Moreover, the controller switching logic should be exclusively designed to fit the characteristics of the candidate controllers. 

\begin{tcolorbox}[size=title, colback=white]
{\textbf{Answer to RQ3:} Building a hybrid controller which strategically switches between AI controllers and traditional controllers is a promising direction and can significantly improve the performance in our evaluated cases. For three types of hybrid controllers we explored, a scenario-dependent approach outperforms the other two in most of the cases.

}
\end{tcolorbox}

\section{Discussions, Future Directions and Threats to Validity}

\myparagraph{Discussions}
According to our insights from RQ1, DRL controllers may fall short of handling multiple control outputs or balancing multiple requirements simultaneously, compared to their traditional counterparts. Indeed, the design of reward function may be too complicated to compensate the balance among different requirements, since rewarding only a portion of the requirements may overshadow others.
In industry, it is a common scenario to handle multiple requirements simultaneously. Therefore, applying DRL-based AI controllers in these cases requires further research.

Based on our evaluation of existing falsification algorithms, we find that falsification may not work effectively for AI-enabled CPS. AI controllers have their specific structure and unique decision logic, which is quite different from their traditional counterparts. Therefore, taking into account the characteristics of this specific formalism is important for effective testing.

The combination of different types of controllers offers a new direction of improving the safety and performance of the controllers, as demonstrated by our evaluation. Specifically, the scenario-dependent approach outperforms others, showing that strategic combination is necessary to achieve superior performances.


\myparagraph{Future Directions}
Based on the insights from our evaluation, we propose the following three future directions for AI-enabled CPS. 
\begin{compactitem}[$\bullet$]
\item First, there is a need for more research efforts on  benchmarks and empirical studies in this direction. Moreover, more complicated system requirements that reflect industrial standards or demands should also be considered for further evaluation;
\item Second, according to RQ2, existing falsification tools are not fully effective in detecting requirement violations in AI-based CPS. This offers a research opportunity of developing more advanced testing techniques for AI-enabled CPS, e.g., by exploiting the specific structure of neural networks for more effective testing; 
\item Third, besides testing, analysis techniques should be developed to understand the root cause of the violations. To achieve this, more research efforts on fault localization and repair are necessary.
\end{compactitem}

\myparagraph{Threats to Validity} In terms of {\em construct validity}, one potential threat is that the evaluation metrics may not fully describe the performance of controllers. To mitigate this threat, we used five evaluation metrics and two falsification tools to comprehensively measure and analyze the performance and reliability of CPS in our benchmark. 
In terms of {\em internal validity}, one potential threat is that the behavior of a CPS can vary when using different environment parameters. To mitigate this threat, we chose to use the same parameters as described in the documentation of each CPS to keep consistency. Further, we confirmed that our simulation results are consistent with the source descriptions and demos. 
In terms of {\em external validity}, one potential threat is that our analysis results may not be generalized to other CPS. To mitigate this threat, we tried our best to collect a diverse set of CPS with different functionalities, system environments, and control tasks. 

\section{Related Work}

\myparagraph{CPS Benchmarks} As mentioned in~\S{}\ref{sec:benchmarks}, collecting benchmarks of CPS is challenging. An annual workshop, namely ARCH, 
aims to mitigate this problem by bringing together CPS benchmarks and holding competitions\footnote{\url{https://cps-vo.org/group/ARCH/FriendlyCompetition}} for different research topics. 
The most relevant competitions to this paper are \emph{Artificial Intelligence and Neural Network Control Systems}~\cite{johnson2020arch} and \emph{Falsification}~\cite{ernst2020arch}. However, the benchmark in~\cite{ernst2020arch} only includes traditional CPS rather than AI-enabled CPS. While the benchmark in~\cite{johnson2020arch} includes AI-enabled CPS, their benchmark includes less and simpler CPS such as \emph{Cart-Pole}, which are not from industrial application domains. 

\myparagraph{AI Controllers for CPS}
Duan \etal~\cite{duan2016benchmarking} proposed a benchmark on continuous control tasks. However, this benchmark involves game scenarios only such as \emph{Cart-Pole} and \emph{Inverted Pendulum}, rather than complex real-world environments.
%
%
Besides DRL, FNN also has been used in designing a tracking controller for a robot manipulator~\cite{braganza2007neural}; however, the FNN controller is a subsystem which can only be used to compensate a feedback controller.

\myparagraph{CPS Testing and Verification} 
Currently, most of the research efforts are devoted to formal verification of such systems, since it can give rigorous proofs on their safety. For example, reachability analysis~\cite{tran2019star, xiang2018output, huang2019reachnn} has been extensively studied and considered as one of the most effective 
ways to verify AI controllers. 
The other line of formal verification of AI controllers is based on constraint solving, such as DLV~\cite{huang2017safety}, \etc.
Due to the intrinsic scalability problem of verification, their evaluations are usually on simple benchmarks.
Falsification is considered as a method that suffers much less from the scalability issue than verification, and this is confirmed by the empirical study~\cite{nejati2019evaluating}, in which they compared the effectiveness of model checking and testing on CPS. However, existing falsification research mostly focuses on CPS with traditional controllers, and does not consider the specific structure of neural networks. 

\section{Conclusion}
This paper presents a public benchmark of AI-enabled CPS in various domains, which can serve as a fundamental evaluation and testing framework for enhancing the understanding and development of AI-enabled CPS. Based on this benchmark, we collected a series of evaluation metrics and measured the performance and reliability of state-of-the-art deep reinforcement learning (DRL) controllers on various types of CPS. Our 
findings reveal some strengths and weaknesses of AI-enabled CPS and highlights an opportunity of strategically combining AI-enabled CPS with traditional CPS. 
Furthermore, our analysis of two widely used falsification techniques on AI-enabled CPS motivates further improvement of these techniques to account for the unique characteristics of AI controllers, in order to build safe and reliable CPS in the age of AI.

\section*{acknowledgment}
This work is supported in part by grant of Future Energy Systems, Canada CIFAR AI Program, NSERC Discovery Grant of Canada, as well as JSPS KAKENHI Grant No.JP20H04168, JST-Mirai Program Grant No.JPMJMI20B8, and JST SPRING Grant No. JPMJSP2136.
\bibliographystyle{ACM-Reference-Format}
\bibliography{short_ref}










\end{document}